\documentclass[sigconf, screen]{acmart}

\AtBeginDocument{%
  \providecommand\BibTeX{{%
    \normalfont B\kern-0.5em{\scshape i\kern-0.25em b}\kern-0.8em\TeX}}}

\setcopyright{iw3c2w3}
\copyrightyear{2021}
\acmYear{2021}
\acmDOI{10.1145/1122445.1122456}

\acmConference[The Web Conference 2021]{The World Wide Web Conference 2021}{April 19-23, 2021}{Ljubljana, Slovenia}
\acmBooktitle{WWW '21: The World Wide Web Conference 2021,
  April 19--23, 2021, Ljubljana, Slovenia}

\usepackage{booktabs}

\usepackage{hyperref}
\usepackage{amsmath,amsfonts}
\usepackage{algorithm}
\usepackage[noend]{algpseudocode}

\usepackage{graphicx}
\usepackage[switch]{lineno} 
\usepackage{multirow}
\usepackage{enumitem}

\begin{document}

\title[RetaGNN: Relational Temporal Attentive GNNs for Holistic Sequential Recommendation]{RetaGNN: Relational Temporal Attentive Graph Neural Networks for Holistic Sequential Recommendation}


\author{Cheng Hsu}
\affiliation{%
  \institution{Institute of Data Science\\
  National Cheng Kung University\\
  Tainan, Taiwan
  }
}
\email{hsuchengmath@gmail.com}

\author{Cheng-Te Li}
\affiliation{%
  \institution{Institute of Data Science\\
  National Cheng Kung University\\
  Tainan, Taiwan
  }
}
\email{chengte@ncku.edu.tw}

\renewcommand{\shortauthors}{Hsu and Li}

\begin{abstract}
Sequential recommendation (SR) is to accurately recommend a list of items for a user based on her current accessed ones. While new-coming users continuously arrive in the real world, one crucial task is to have 
\textit{inductive} SR that can produce embeddings of users and items without re-training. Given user-item interactions can be extremely sparse, another critical task is to have \textit{transferable} SR that can transfer the knowledge derived from one domain with rich data to another domain. In this work, we aim to present the \textit{holistic SR} that simultaneously accommodates conventional, inductive, and transferable settings. We propose a novel deep learning-based model, \textit{Relational Temporal Attentive Graph Neural Networks} (RetaGNN), for holistic SR. 
The main idea of RetaGNN is three-fold. First, to have inductive and transferable capabilities, we train a \textit{relational attentive GNN} on the local subgraph extracted from a user-item pair, in which the learnable weight matrices are on various relations among users, items, and attributes, rather than nodes or edges. Second, long-term and short-term temporal patterns of user preferences are encoded by a proposed \textit{sequential self-attention} mechanism. Third, a \textit{relation-aware} regularization term is devised for better training of RetaGNN.
Experiments conducted on MovieLens, Instagram, and Book-Crossing datasets exhibit that RetaGNN can outperform state-of-the-art methods under conventional, inductive, and transferable settings. The derived attention weights also bring model explainability.

\end{abstract}



\maketitle

\section{introduction} 
\label{sec:introduction}
Sequential recommendation (SR) is one of the crucial research lines in recommender systems (RS)~\cite{recsurv19}. SR considers the chronological order of user-item interactions, and models the correlation between a user's recent successively interacted items and the choices of the next ones. Given a sequence of items recently accessed by a user, the goal of SR is to learn the sequential preference of her so that future items can be recommended accurately. The SR task differs from conventional RS. RS aims to capture user's global preferences on items~\cite{mfhe16,plsr18,ngcf}. SR targets at learning the sequential patterns of user-item interactions based on the recent sequence of items. In other words, SR requires the modeling of long-term and short-term interests and intents of users~\cite{caser18,hgn19,ma2020memory} in predicting next items.

One of the mainstream approaches to RS is matrix factorization (MF). MF generates the embeddings of users and items in a transductive manner, which refers to utilizing the rich user-item interactions during training. However, when there arrive new users or unseen items that have never been interacted with existing ones, their embeddings cannot be learned. Techniques on inductive matrix completion (IMC)~\cite{imc1,imc2} deal with such an issue by leveraging content information of users and items. For example, user attributes or item tags are used to serve as the bridge between new users/items and existing ones. PinSage~\cite{pinsage} further resorts to visual and textual content associated with items for inductive RS. 
Factorized exchangeable autoencoder (FEAE)~\cite{feae} alternatively develops exchangeable and permutation-equivariant matrix operations to perform inductive RS without using item content. However, FEAE cannot be scalable to large user-item matrices.
A recent advance IGMC~\cite{igmc} presents a graph neural network (GNN) based IMC model that relies on item content and requires only local user-item subgraphs, which leads to both inductive and scalable RS.

While some of the existing RS methods have been capable of inductive learning, state-of-the-art sequential recommendation models, such as HGN~\cite{hgn19}, HAM~\cite{peng2020ham} and MA-GNN~\cite{ma2020memory}, are still transductive. To the best of our knowledge, the task of inductive SR is not formally explored yet. This work aims at inventing an effective inductive SR model. In addition, we think that existing SR models can be further improved since two factors had not been considered yet. The first is the modeling of high-order user-item interactions in long and short terms of the given sequence. The sequential evolution of multi-hop collaborative neighbors of a user in the interaction graph can reveal how user preferences change over time.
The second is the temporal patterns in the derived representations of sequential items. The adoption of the next items can be influenced by recent items with different weighting contributions.

\begin{figure*}[!t]
  \centering
  \includegraphics[width=\textwidth]{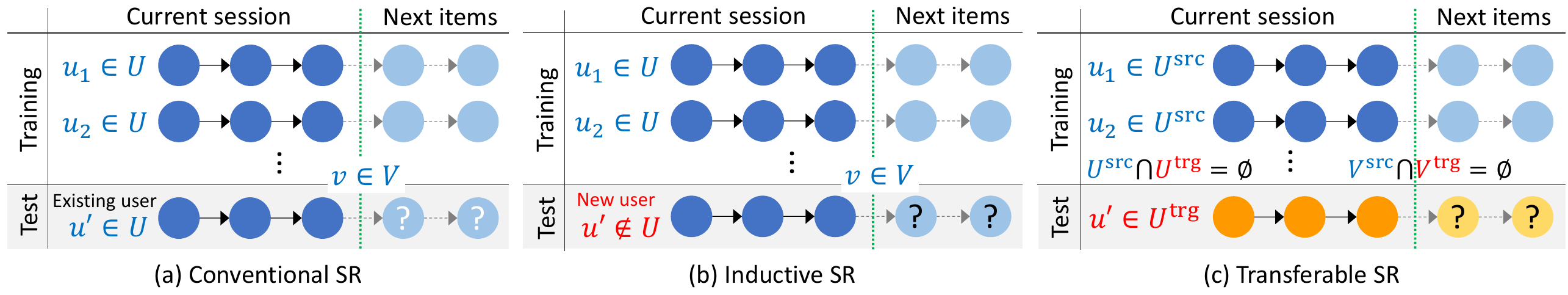}
  \caption{The present work: holistic sequential recommendation with conventional, inductive, and transferable settings.}
  \vspace{-5pt}
   \label{fig:holisr}
\end{figure*}

In this paper, we propose a novel deep learning-based model, \textit{\underline{RE}lational \underline{T}emporal \underline{A}ttentive \underline{G}raph \underline{N}eural \underline{N}etwork} (RetaGNN), for sequential recommendation. In a particular domain with fixed sets of users, items, and their interactions, given a sequence of recent interacted items for a user, our main goal is three-fold, as illustrated in Figure~\ref{fig:holisr}. The first is \textit{conventional SR}: to accurately recommend the next items. The second is \textit{inductive SR}: to recommend the next items to new-coming users that are not contained in the existing user set. The third is \textit{transferable SR}: to recommend the next items for a given user's item sequence, in which both users and items belong to another domain. That said, our goal is to have a \textit{holistic SR} model that can predict the next items under conventional, inductive, and transferable settings. It is challenging to simultaneously deal with three SR settings since the learnable parameters of a model should NOT be attached to a specific node (for inductive) or a particular dataset (for transferable). That said, the model needs to capture common knowledge shared across seen/unseen nodes and various datasets. 

To achieve the abovementioned SR goals, the proposed RetaGNN has four major ideas. First, RetaGNN is built upon \textit{individual user-item pair's local graph patterns}, in which the graph involves relations among users, items, and attributes. We extract the local subgraph surrounded by the given user-item pair from the given existing sequential user-item interactions, and learn to map such a subgraph to a score of their interaction. Second, to have both inductive and transferable capabilities, we present a \textit{Relational Attentive GNN} (RA-GNN) layer to model high-order user-item interactions in the sequential context. RA-GNN performs the message passing and neighborhood aggregation mechanisms in RetaGNN by training learnable weight matrices on various relations, rather than on nodes (e.g., GAT~\cite{gat18} and NGCF~\cite{ngcf}), in the graph. Based on such two ideas, as long as we can obtain the local subgraph regarding the given user-item pair, no matter whether the user is seen or not, the relation weights can be applied to generate user and item embeddings in both inductive and transferable settings. Third, we propose a \textit{Sequential Self-Attention} (SSA) layer to encode the temporal patterns from the RA-GNN generated sequential item embeddings. Last, we devise a \textit{relation-aware regularization} term into the loss function so that learnable parameters associated with relations in RetaGNN can be better trained.

The contributions of this work are summarized as follows.
\begin{itemize}[leftmargin=*]
\item We are the first to holistically tackle the sequential recommendation task that simultaneously accommodates conventional, inductive, and transferable settings.
\item We present a novel Relational Temporal Attentive Graph Neural Network (RetaGNN)
model to achieve the goal. The main idea is to learn the mapping from a local graph of the given user-item pair to their interaction score, and to train the learnable relation weight matrices. 
\item We conduct the experiments on MovieLens, Instagram, and Book-Crossing datasets, and the results show that RetaGNN can outperform state-of-the-art SR models and inductive RS models under conventional, inductive, and transferable settings. The derived sequential attention weights also bring model explainability.
\end{itemize}

This paper is organized as follows. Section~\ref{sec:related_work} reviews relevant studies, and Section~\ref{sec:problem} describes the problem statement. We give the technical details of our RetaGNN model in Section~\ref{sec:methodology}, followed by presenting the experimental results in Section~\ref{sec:experiments}. We conclude this work in Section~\ref{sec:conclusion_and_future_work}.


\section{related work} 
\label{sec:related_work}
The review of relevant studies is divided into three parts: sequential recommendation models, inductive recommendation systems, and knowledge graph-enhanced recommendation systems. We provide a detailed discussion for each part in the following.

\textbf{SR Models.} In deep SR models, recurrent neural networks~\cite{gru4rec16,rrn17,rnnsr18} and convolutional neural networks~\cite{caser18} are used to extract long-term and short-term sequential features. SASRec~\cite{sasrec18} is a self-attentive model that can identify the most significant items for prediction. MARank~\cite{marank19} is a multi-order attentive ranking model that unifies both individual- and union-level item-item interaction into preference inference from multiple views. NextItNet~\cite{nextitnet19} is a dilated convolution-based generative method to learn long-range dependencies in the item sequence. JODIE~\cite{jodie19} is a coupled recurrent neural network model that jointly learns the embedding trajectories of users and items from a sequence of temporal interactions. 
SR-GNN~\cite{srgnn} is a GNN-based SR model that learns item embeddings by applying GNN to the graph from item sequences.
HGN~\cite{hgn19} is a hierarchical gating neural network that adopts a feature gating and instance gating to determine what item features should be used for recommendation. HAM~\cite{peng2020ham} further models sequential and multi-order user-item association patterns for SR. The state-of-the-art MA-GNN~\cite{ma2020memory} leverages graph neural networks to model the item context, along with a shared memory network to capture item-item dependencies. Although past studies have brought some success, the inductive learning in SR does not be investigated yet. To the best of our knowledge, our work is the first attempt to inductive SR.

\begin{figure*}[!t]
  \centering
  \includegraphics[width=1.0\textwidth]{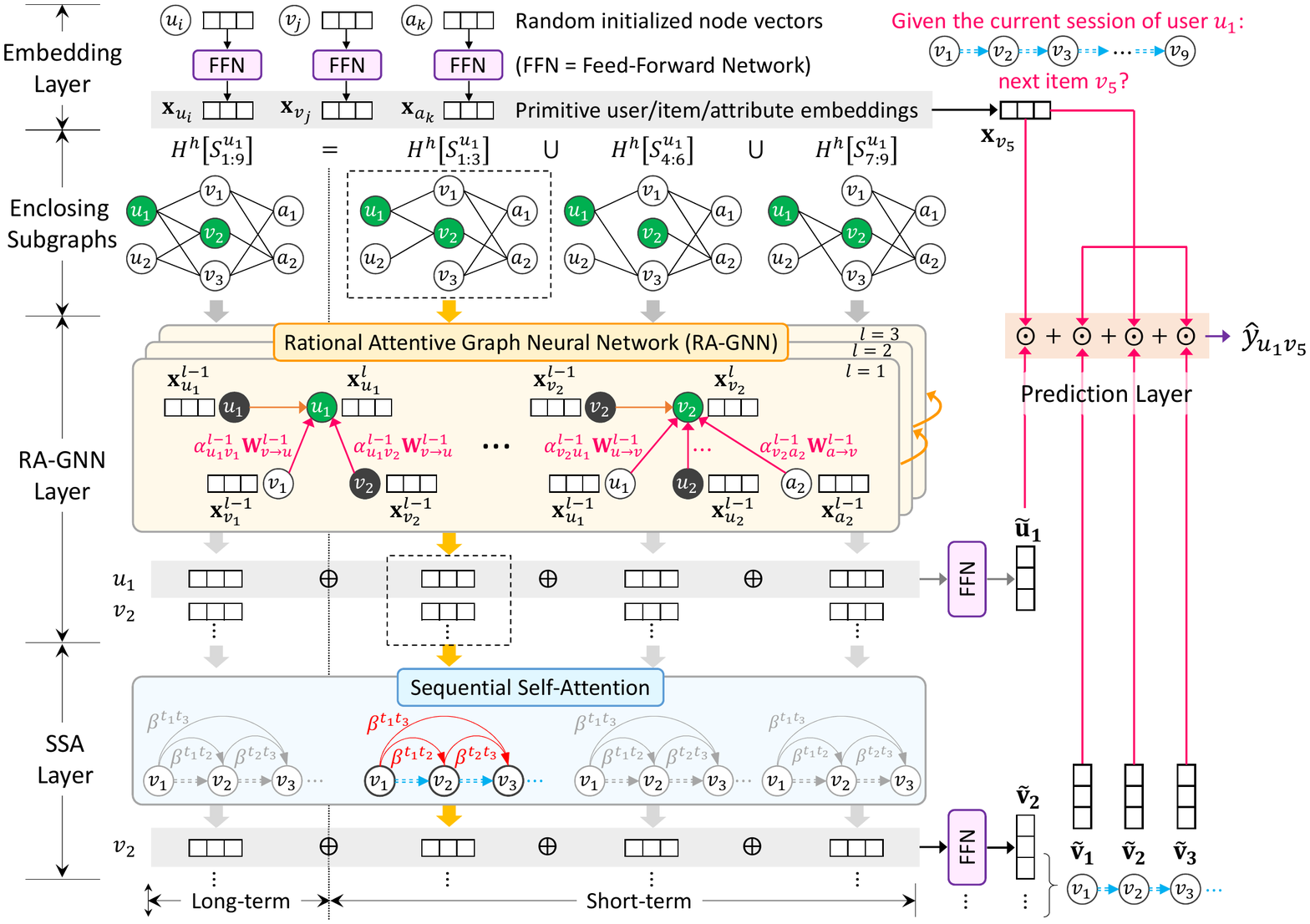}
  \caption{Model overview of the proposed RetaGNN for holistic sequential recommendation. The input session $\mathcal{S}^u_{1:9}=(v_1,v_2,...,v_9)$ is created by user $u_1$, and the next item being predicted is $v_5$. We utilize user $u_1$ and item $v_2$ to be the target pair $(u_1,v_2)$ to extract enclosing subgraphs and perform RA-GNN. $\oplus$ means concatenation, and $\odot$ indicates dot product.}
   \label{fig:overview}
  \vspace{-5pt}
\end{figure*}

\textbf{Inductive RS.}
Recent advances of RS, including GCMC~\cite{berg2017graph}, HGAN~\cite{wang2019heterogeneous}, NGCF~\cite{ngcf}, and LightGCN~\cite{lightgcn}, employ graphs to depict various user-item interactions, together with graph representation learning, to obtain promising performance. However, not many eyes put on inductive RS. 
Inductive matrix completion (IMC)~\cite{imc1,imc2} first utilizes content information such as user attributes and item categories for inductive RS. PinSage~\cite{pinsage} and TransGRec~\cite{transgrec} further consider rich multimedia content associated with items for inductive learning. FEAE~\cite{feae} does not rely on content, but presents a factorized exchangeable autoencoder with permutation-equivariant matrix operations to make RS inductive. Although GraphSage~\cite{graphsage} and GraphSAINT~\cite{graphsaint} can perform inductive graph representation learning, they are applicable to simple graphs, rather than the bipartite graphs of user-item interactions in RS. The state-of-the-art is the graph neural network-based inductive matrix completion (IGMC)~\cite{igmc}. IGMC represents the collaborative neighbors of a user-item pair as a local subgraph, and performs graph-level GNN to map the subgraph to an interaction probability of that pair. The IGMC embedding generation is inductive because it relies on only the local subgraph of the user-item pair. That said, any new-coming users can be linked to existing items even they never access items. Our work aims to extend the idea of IGMC to SR.

\textbf{KG-enhanced RS.} Knowledge graph (KG) embedding~\cite{kgesur17} brings auxiliary features depicting the correlation between items through their metadata and attributes. KGs are leveraged in various ways in RS, including propagating user preferences over knowledge entities by RippleNet~\cite{rpnet18}, multi-task learning with KG Embedding by MKR~\cite{mkr19}, applying graph attention on a user-item-attribute graph by KGAT~\cite{kgat19}, adopting LSTM to model sequential dependencies of entities and relations~\cite{kprn19}, and integrating induction of explainable rules from KG by RuleRec~\cite{exrule19}. Although these successfully apply KG to RS, it remains limited in utilizing KG for inductive and transferable SR, which can be a potential future extension of our work. Heterogeneous information network (HIN) embedding approaches, such as HetGNN~\cite{hetgnn19} and GATNE~\cite{gatne19}, can also learn transductive user representations by considering the interactions and correlation between users and diverse entities. They can be directly used for RS, rather than holistic SR tasks in our work.


\section{Problem Statement} 
\label{sec:problem}
In the context of recommender systems, we have a set of $M$ users $\mathcal{U}$ = $\{u_{1},u_{2},...,u_{M}\}$ and a set of $N$ items $\mathcal{V}$ = $\{v_{1},v_{2},...,v_{N}\}$. The matrix of user-item interaction is denoted by $Y\in \mathbb{R}^{M\times N}$ based on implicit feedback from users, in which $y_{uv}$ = 1 indicates user $u$ has interacted with item $v$; otherwise, $y_{uv}$ = 0. A user $u$ can sequentially interact with a number of items at different time steps. A sequence of consecutive interacted items is termed a \textit{session}. Given a user $u$, we denote one of her sessions as $S^{u} = (s^{u}_{1},s^{u}_{2},...,s^{u}_{L})$, where $L=|S^{u}|$ is the length of session $S^{u}$, and $s^{u}_{i} \in \mathcal{V}$ is an item index that user has interacted with. We also denote the set of items that user $u$ interacted with as $V_u$. Let $\mathcal{A}$ be the universe set of item attributes and $\mathcal{A} = \{ a_{1},a_{2},...,a_{k} \}$, where $k$ is the number of total item attribute values. We can denote the attribute value set of item $v$ as $\mathcal{A}_v$.
With these notations, the \textit{holistic sequential recommendation} problem can be defined from three aspects. 

(1) \textit{Conventional Sequential Recommendation} (CSR): given the earlier session $S^{u}_{1:t} (t<L)$ of every user $u \in \mathcal{U}$, we aim to recommend a list of items from item set $\mathcal{V}$ to each user. In other words, the goal is to predict whether user $u$ will interact with item $v \in \mathcal{V}$ after time $t$ (i.e., whether the items in the recommended item list will appear in the ground truth $S^u_{t:L}$).

(2) \textit{Inductive Sequential Recommendation} (ISR): given the earlier session $S^{u}_{1:t} (t<L)$ of every user $u \in \mathcal{U}^{\bullet}$, we aim to recommend a list of items from item set $\mathcal{V}$ to each user $u' \in \mathcal{U}^{\circ}$, where $\mathcal{U}^{\bullet}$ is the seen user set, $\mathcal{U}^{\circ}$ is the unseen user set (users do not appear at the training stage), and $\mathcal{U}^{\bullet} \cap \mathcal{U}^{\circ} = \emptyset$. That said, the goal is to return a list of items (from $\mathcal{V}$) that an unseen user $u'\in \mathcal{U}^{\circ}$ will interact with in the near future.

(3) \textit{Transferable Sequential Recommendation} (TSR): given the earlier session $S^{u}_{1:t} (t<L)$ of every user $u \in \mathcal{U}^{\text{src}}$ in the source domain, we aim at producing a sequential recommendation model with transferable parameters $\Theta$ that can recommend a list of items from item set $\mathcal{V}^{\text{src}}$ for each user $v \in \mathcal{U}^{\text{src}}$. Then by applying the transferable parameters $\Theta$ to the target domain with user set $\mathcal{U}^{\text{trg}}$ and item set $\mathcal{V}^{\text{trg}}$, where $\mathcal{U}^{\text{src}} \cap \mathcal{U}^{\text{trg}} = \emptyset$ and $\mathcal{V}^{\text{src}} \cap \mathcal{V}^{\text{trg}} = \emptyset$, we can accurately recommend a list of items (from $\mathcal{V}^{\text{trg}}$) to every user $u'\in \mathcal{U}^{\text{trg}}$.


\section{methodology} 
\label{sec:methodology}
We present the overview of the proposed RetaGNN model in Figure~\ref{fig:overview}. RetaGNN consists of five phases. First, we utilize a one-hidden-layer feed-forward network (FFN) to generate the primitive embeddings of users, items, and attributes. Second, we extract the long-term and short-term $h$-hop enclosing subgraphs for every target pair compiled by pairing a user and each of her interacted items at different time frames. Third, a relation attentive graph neural network (RA-GNN) layer is created to learn the representations of users and items, which encodes the sequential high-order user-item interactions, in every enclosing subgraphs. The inductive and transferable learnable parameters can be obtained in this phase. Fourth, we devise a sequential self-attention (SSA) layer to model the temporal patterns of user preferences, and item embeddings are updated here. Last, by leveraging the primitive embedding of an item being predicted, along with the sequential item embeddings, the prediction layer produces the scored results.

\begin{figure}[!t]
  \centering
  \includegraphics[width=\linewidth]{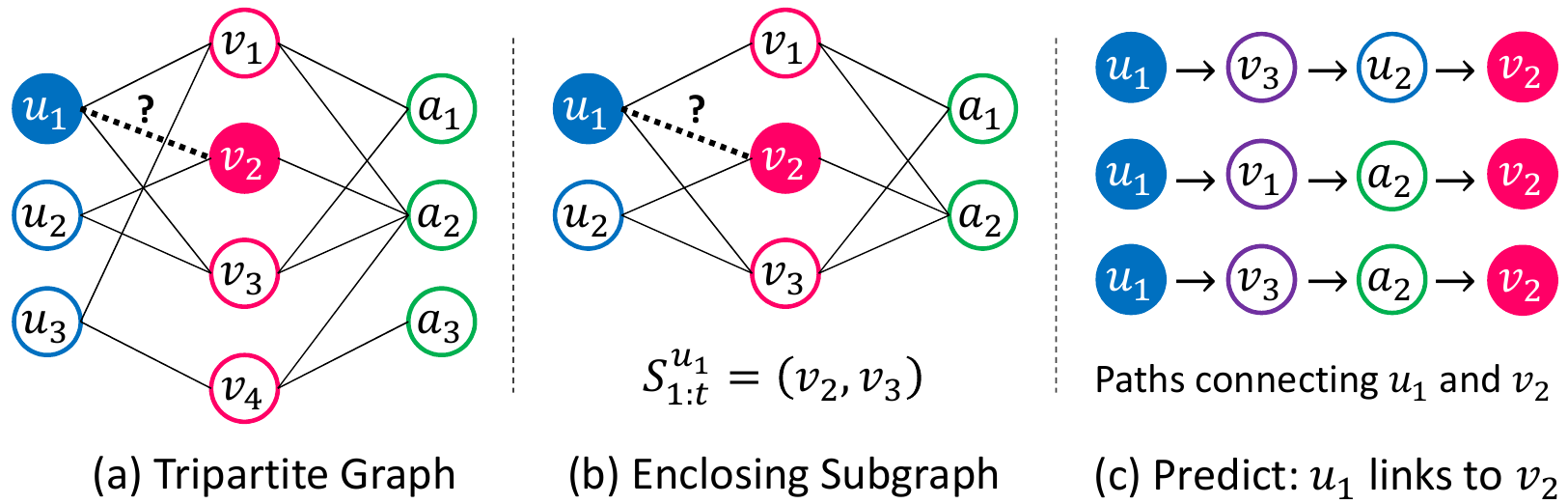}
  \caption{A toy example on the tripartite graph, an enclosing subgraph, and paths used to predict the interaction between user $u_1$ and item $v_2$.}
  \label{fig:toy}
\end{figure}

\subsection{Primitive Embedding Layer}
We first consider to randomly initialize the representation vectors of all users, items, and attribute values. The reason is for inductive and transferable learning. The randomly-initialized ``fixed-size'' vectors on nodes allow us to update the embeddings of both new-coming unseen nodes (for inductive) and cross-data nodes (for transferable) under the same set of learned model weights. The reason is that we learn model weights on \textit{directional edge relations} in the constructed graph, which is independent of nodes and datasets. Hence, RetaGNN can project the randomly-initialized vectors of new-coming and cross-data nodes into the same embedding space to achieve the inductive and transferable effects.

By feeding randomly-initial vectors into an embedding layer, i.e., one-hidden-layer feed forward network (FFN), we can generate a low-dimensional real-value dense vector for every user, item, and attribute value. We term such dense vectors \textit{primitive embeddings}. Let the embeddings be $\mathbf{X}\in\mathbb{R}^{q\times d}$, where $d$ is the embedding dimension, and $q$ is the sum of numbers of users, items, and attribute values in the training data. Given a session $\mathcal{S}^u_{1:t}$ of a particular user $u$, its corresponding embedding matrix can be represented by: $\mathbf{X}_{\mathcal{S}^u_{1:t}}=[\mathbf{x}_{1} \cdots \mathbf{x}_{j} \cdots \mathbf{x}_{t}]$,
where $\mathbf{X}_{\mathcal{S}^u_{1:t}}\in\mathbb{R}^{t\times d}$, and $\mathbf{x}_j\in \mathbb{R}^d$ is the primitive embedding of the $j$-th item in $\mathcal{S}^u_{1:t}$ and can be retrieved from the primitive embedding matrix $\mathbf{X}$.
Note that here we do not utilize the fixed-length one-hot encodings to initialize the vectors. The reason is that one-hot encoding is not extensible for new-coming users and cross-domain users/items, and thus prohibits new-coming and other-domain users from producing their primitive embeddings.

\subsection{User-Item-Attribute Tripartite Graph}
We construct a global tripartite graph $\mathcal{H}$ to represent the relationships among users, items, and item attributes. Let the tripartite graph be $\mathcal{H}=(\mathcal{N},\mathcal{E})$, where the node set $\mathcal{N}$ is the union of user set $\mathcal{U}$, item set $\mathcal{V}$, and attribute value set $\mathcal{A}$, i.e., $\mathcal{N}=\mathcal{U} \cup \mathcal{V} \cup \mathcal{A}$. The edge set $\mathcal{E} = \mathcal{E}^{\mathcal{U}\mathcal{V}} \cup \mathcal{E}^{\mathcal{V}\mathcal{A}}$, where $\mathcal{E}^{\mathcal{U}\mathcal{V}}$ and $\mathcal{E}^{\mathcal{V}\mathcal{A}}$ are the sets of edges connecting users with items, and connecting items with their attribute values, respectively. In other words, there are no edges between same-type nodes, and no edges between users and item attribute values. A toy example of the tripartite graph is given in Figure~\ref{fig:toy}(a). Note that the tripartite graph $\mathcal{H}$ is constructed based on different training sessions, i.e., long-term and short-term ones described in the following subsections.

The purpose of the tripartite graph is three-fold. First, the paths via user-item interactions can reflect the effect of collaborative filtering~\cite{ncf,ngcf,defm20}. For example in Figure~\ref{fig:toy}(c), to determine whether to recommend item $v_2$ to user $u_1$, the path $(u_1,v_3,u_2,v_2)$ can reflect that users $u_1$ and $u_2$ share similar taste based on item $v_3$, we can recommend item $v_2$ to $u_1$ because $u_3$ also likes $v_2$. Second, the paths via item-attribute interaction can depict the correlation between items, which can to some degree bring the effect of content-based filtering. For example in Figure~\ref{fig:toy}(c), both paths $(u_1,v_1,a_2,v_2)$ and $(u_1,v_3,a_2,v_2)$ imply that we can recommend $v_2$ to $u_1$ because items $v_2$ share the same attribute $a_2$ with items $v_1$ and $v_3$, which were liked by $u_1$. Third, with the tripartite graph, our model is allowed to be capable of inductive and transfer learning, i.e., dealing with new-coming users, who can be put in the graph so that we can obtain relevant paths to connect them with existing users/items and accordingly make prediction. We will elaborate the details in the following.

There are three types of nodes in the tripartite graph. To encode more semantics into paths that depict the relationships between users and items, we think that the edges from one node type to another can have different meanings. Edges from a user $u_i$ to an item $v_j$, from $v_j$ to $u_i$, from $v_j$ to an attribute $a_k$, and from $a_k$ to $v_j$, represent: $u_i$ \textit{likes} $v_j$, $v_j$ \textit{is adopted by} $u_i$, $v_j$ \textit{has attribute} $a_k$, and $a_k$ \textit{is possessed by} $v_j$, respectively. Hence, we consider them as four different \textit{relations}. Let $\mathcal{R}$ denote the relation set, and $|\mathcal{R}|=4$. We define a mapping function $R(n_i,n_j)$ that can map two nodes $n_i,n_j\in \mathcal{N}$ into their corresponding relation $r\in \mathcal{R}$.

\subsection{Extracting Enclosing Subgraphs}
We first prepare the set of positive user-item pairs by pairing a user $u$ with each of her interacted item $v$ in session $\mathcal{S}^{u}_{a:b}$. Given a user $u$ and an item $v$, i.e., the target pair $(u,v)$, and the tripartite graph $\mathcal{H}_{a:b}$ constructed from all sessions $\mathcal{S}^{u}_{a:b}$ ($a<b$) $\forall u\in\mathcal{U}$, we extend the IGMC's subgraph modeling~\cite{igmc} to extract an enclosing subgraph $\mathcal{H}^{h}_{a:b}[u,v]$ from the tripartite graph $\mathcal{H}_{a:b}$, where $1\leq a<b\leq t$. To depict the high-order relationships among users, items, and attributes, we utilize a hop number $h$ to determine the size of the extracted enclosing subgraph. We will examine how $h$ affects the performance in the experiments. A higher $h$ value allows us to depict higher-order user-item-attribute relationships, but it also brings higher computational complexity. Algorithm~\ref{alg:extsub} presents the breath-first search (BFS) strategy, which is centered at the target pair of user $u$ and item $v$ in tripartite graph $\mathcal{H}_{a:b}$, to extract the $h$-hop enclosing subgraphs. Each extracted subgraph $\mathcal{H}^{h}_{a:b}[u,v]$ is used to train the GNN model, and to generate the representations of users and items for predicting next sequential ones for user $u$.

\begin{algorithm}[!t]
\caption{Extracting Enclosing Tripartite Subgraphs}
\label{alg:extsub}
\begin{flushleft}
\textbf{Input:} hop number $h$, the target pair of user $u$ and item $v$, the tripartite graph $\mathcal{H}_{a:b}$ constructed from all sessions $S^{u}_{a:b}$ ($a<b$) of all users $u\in\mathcal{U}$, the universe set of users $\mathcal{U}$ and the universe set of attributes $\mathcal{A}$ \\
\textbf{Output:} the $h$-hop extracted enclosing subgraph $\mathcal{H}^{h}_{a:b}[u,v]$
\end{flushleft}
\begin{algorithmic}[1]
\State $U = U_{f}=\{u\}$, $V = V_{f}=\{v\}$, $A = A_f=\emptyset$ 
\For  {$i=1,2,...,h$}
\State $U'_{f} = \{u_{i}: u_{i} \sim V_{f} \} \setminus (U \cup \mathcal{A})$ 
\State $V'_{f} = (\{v_{i}: v_{i} \sim U_{f} \} \setminus V) \cup (\{a_{i}: a_{i} \sim A_{f} \} \setminus V)$
\State $A'_{f} = \{a_{i}: a_{i} \sim V_{f} \} \setminus (A \cup \mathcal{U})$
\State $U = U'_{f}$, $V = V'_{f}$, $A = A'_{f}$
\State $U$ = $U\cup U_{f}$, $V$ = $V\cup V_{f}$, $A$ = $A\cup A_{f}$
\EndFor
\State Let $\mathcal{H}^{h}_{a:b}[u,v]$ be the vertex-induced subgraph from $\mathcal{H}_{a:b}$ usig vertex sets $U$ and $V$\\
\Return $\mathcal{H}^{h}_{a:b}[u,v]$
\end{algorithmic}
\begin{flushleft}
\noindent Note: $\{ u_{i} : u_{i} \sim V_{f} \}$ is the set of nodes that are adjacent to at least one node in $V_{f}$ with any edge type.
\end{flushleft}
\vspace{-5pt}
\end{algorithm}

\subsection{Relational Attentive GNN Layer}
To learn the feature representation of each user and item node in extracted subgraphs, we devise Relational Attentive Graph Neural Network (RA-GNN). The input is an enclosing subgraph, and the output is the embedding of each node. The RA-GNN consists of two parts. One is the \textit{relational attention} mechanism, and the other is the message passing between nodes. In the relation attention mechanism, we consider that different relations have various contributions to their incident nodes in the subgraph. Learning the attention weights of relations with respect to different users needs to jointly model both user preferences and item correlation. For example, for a particular user, her preference can be jointly reflected by other users who have the same taste as her, and also by items possessing the same attribute values as her existing interacted ones, along with different weights. In addition, two items can be correlated with one another in terms of common interacted users or common attributes, with various weights. In the message passing, we aim at depicting each user and item using their high-order paths connected with other items, users, and attribute values, as illustrated in Figure~\ref{fig:toy}(c). In short, the proposed RA-GNN will learn the rich pathway patterns with different attention weights on relations to represent each user and item in every extracted subgraph.

Let the initial vector of each node $v_i\in \mathcal{G}$ be $\mathbf{x}^0_i\in\mathbb{R}^d$ obtained from the primitive embedding matrix $\mathbf{X}$, where $\mathcal{G}=\mathcal{H}^{h}_{a:b}[u,v]$ is an enclosing subgraph and $d$ is the dimension of RA-GNN embedding vector. Also let $\Gamma_r(v_i)$ be the set of incident neighbors of node $v_i$ via relation (i.e., directed edge type) $r\in\mathcal{R}$ in graph $\mathcal{G}$. To have the updated embedding of node $v_i$ at the $(l+1)$\textsuperscript{th} layer, denoted by $\mathbf{x}^{l+1}_i$, from $\mathbf{x}^{l}_i$, a two-step method is devised. The first step is the relational attention mechanism that aggregates the embedding of all $v_i$'s neighbors $v_j$ by the relational attention weights $\alpha^{l}_{ij}\mathbf{W}^{l}_{r}$. The second step is to have it combined with $\mathbf{x}^{l}_i$. Such two steps can be depicted by the following equation:
\begin{equation}
\label{eq:ragnn}
\mathbf{x}^{l+1}_{i}=\mathbf{W}^{l}_{o}x^{l}_{i}+\sum_{r\in\mathcal{R}}\sum_{v_j\in \Gamma_r(v_i)}\alpha^{l}_{ij}\mathbf{W}^{l}_{r}\mathbf{x}^{l}_{j},
\end{equation}
where $\mathbf{W}^{l}_{o} \in \mathbb{R}^{1\times d}$ and $\mathbf{W}^{l}_{r} \in \mathbb{R}^{1\times d}$ are matrices of learning parameters, and $\mathbf{x}^{l}_{i}$ is the embedding vector of node $v_i$ at layer $l$. The relation $r\in\mathcal{R}$ is one of the four relations between nodes $v_i$ and $v_j$. The attention weight $\alpha^{l}_{ij}$ between nodes $v_i$ and $v_j$ can be obtained by first transforming their corresponding embedding via $\mathbf{W}^l_o$ and $\mathbf{W}^l_r$, concatenating them, having another transformation via a weight vector $\mathbf{a}$, passing a non-linear activation function, and last normalizing by the softmax function. The generation of attention weight $\alpha^{l}_{ij}$ is given by:
\begin{equation}
\label{eq:relatt}
\alpha^{l}_{ij}=softmax\left(\rho\left(\mathbf{a}^{\top}\left[\mathbf{W}^{l}_{o}\mathbf{x}^{l}_{i} \oplus \mathbf{W}^{l}_{r}\mathbf{x}^{l}_{j}\right]\right)\right),
\end{equation}
where $\rho$ denotes the LeakyReLU function, $\mathbf{a} \in \mathbb{R}^{1\times d}$, and $\oplus$ is the concatenation operator.

The design of $\mathbf{W}^{l}_{o}$ and $\mathbf{W}^{l}_{r}$ aims at distinguishing the updating node from its neighbors in both Eq.~\ref{eq:ragnn} and Eq.~\ref{eq:relatt}. The transformation by $\mathbf{W}^{l}_{r}$ also has the effect on distinguishing neighbors with different relations $r$ so that various subgraph patterns can be highlighted to depict various user preferences and item correlation. The equation Eq.~\ref{eq:relatt} that generates attention weight  $\alpha^{l}_{ij}$ is \textit{layer-dependent}. The weight matrices $\mathbf{W}^{l}_{o}$ and $\mathbf{W}^{l}_{r}$ at different layers $l$ lead to different attention weights $\alpha^{l}_{ij}$ between $v_i$ and $v_j$. Such a design of attention mechanism can learn the contributions of different hops away from the updating node (i.e., different layers of graph neural network). We think having layer-dependent relational attention weight is important since every enclosing subgraph is tripartite with four relations. 

In summary, equipped with the layer-wise attention weights $\alpha^{l}_{ij}$ and different layers' distinguishing weight matrices $\mathbf{W}^{l}_{o}$ and $\mathbf{W}^{l}_{r}$, we can learn a variety of enriched subgraph patterns involving different relations among users, items, and attributes. The relational attentive graph neural network can eventually generate effective user and item representations that encode user preferences and item correlation via their high-order relationships for sequential recommendation. For simplicity and follow-up usages, we denote the RA-GNN derived embedding matrices of users, items, and attributes as $\mathbf{U}$, $\mathbf{V}$, and $\mathbf{A}$, respectively.

Eventually, we can have a mapping function $RAG(\cdot)$ to denote the RA-GNN embedding generation, given by: $\bar{\mathbf{H}}$ = $RAG(\mathcal{G})$, where $\mathcal{G}$ is the input graph of RA-GNN, and $\bar{\mathbf{H}}$ is the output matrix of node representations. Given the extracted enclosing subgraph $\mathcal{H}^{h}_{a:b}[u,v]$, we can generate the corresponding matrix of node embeddings $\bar{\mathbf{H}}=\{\mathbf{U}, \mathbf{V}, \mathbf{A} \}$. We also denote the sequential item embedding matrix $\mathbf{V}^u_{a:b}=(\mathbf{v}_a, \mathbf{v}_{a+1}, ..., \mathbf{v}_{b})$ for session $\mathcal{S}^u_{a:b}$, where $\mathbf{v}_a, \mathbf{v}_{a+1}, ..., \mathbf{v}_{b}\in \mathbf{V}$.

\subsection{Sequential Self-Attention}
Since the task is sequential recommendation, we need to learn the sequential correlation between items within the given session $\mathcal{S}^{u}_{a:b}$. We present a sequential self-attention mechanism to generate the representations of items with their temporal information. The input of sequential self-attention is the sequence of item embeddings derived from RA-GNN, denoted as: $\mathbf{V}^u_{a:b}=\left[\mathbf{v}_a,\mathbf{v}_{a+1},...,\mathbf{v}_b\right]$, where $\mathbf{v}_t\in\mathbb{R}^d$, $a\leq t\leq b$, and $d$ is the embedding dimension. The output is the matrix of sequentially interacted item embeddings, denoted as $\mathbf{Z}^u_{a:b}=\left[\mathbf{z}_a,\mathbf{z}_{a+1},...,\mathbf{z}_b\right]$, where $\mathbf{v}_t\in\mathbb{R}^d$. We denote the sequential self-attention mechanism as the $SSA(\cdot)$ function: $\mathbf{Z}^u_{a:b}=SSA(\mathbf{V}^u_{a:b})$.

The main idea of the sequential self-attention is to model how items at different time steps in the current session sequentially influences their future ones. We take advantage of the scaled dot-product attention~\cite{attall17} to generate item embeddings with sequential self-attention, in which queries, keys, and values are the existing item embeddings $\mathbf{V}^u_{a:b}$. We first learn three linear projection matrices, $\mathbf{W}^{\text{que}} \in \mathbb{R}^{d\times d}$, $\mathbf{W}^{\text{key}} \in \mathbb{R}^{d\times d}$, and $\mathbf{W}^{\text{val}} \in \mathbb{R}^{d\times d}$, to transform the queries, keys, and values to their respective spaces. For each $t$\textsuperscript{th} item's embedding $\mathbf{v}_t\in\mathbf{V}^u_{a:b}$ ($a\leq t\leq b$), we make it be attended by all of the items before and including time step $t$. We do not allow $\mathbf{v}_t$ to attend to items at future time steps $t+1, t+2, ..., b$ since the interaction with the $t$\textsuperscript{th} item is not possible to be determined by its future items.

For a sequential item embedding matrix $\mathbf{V}^u_{a:b}$ of session $\mathcal{S}^{u}_{a:b}$, we learn the attention weight matrix $\mathbf{\beta}_{\mathcal{S}^{u}_{a:b}} \in \mathbb{R}^{T\times T}$, where $T=a-b+1$ is the number of time steps in session $\mathcal{S}^u_{a:b}$, based on the temporal order-aware multiplicative computation on the projected query and key matrices. The sequential self-attention is applied to the projected value matrix to generate the output embedding matrix $\mathbf{Z}^u_{a:b}\in\mathbb{R}^{T\times d}$ of items $v\in\mathcal{S}^u_{a:b}$. Specifically, we define the sequential self-attention as follows:
\begin{equation}
\mathbf{Z}^u_{a:b} = \mathbf{\beta}_{\mathcal{S}^{u}_{a:b}}\left(\mathbf{V}^u_{a:b}\mathbf{W}^{val}\right)
\end{equation}
\begin{equation}
\beta^{t_it_j}_{\mathcal{S}^{u}_{a:b}} = \frac{\exp(e^{t_it_j}_{V})}{\sum_{k=1}^{T}\exp(e^{t_it_k}_{V})}
\end{equation}
\begin{equation}
e^{t_it_j}_{V} = \frac{\left(\left(\mathbf{V}^u_{a:b}\mathbf{W}^{que}\right)\left(\mathbf{V}^u_{a:b}\mathbf{W}^{key}\right)^{\top} \right)_{t_it_j}}{\sqrt{d}} + I_{t_it_j},
\end{equation}
where $\mathbf{I}\in\mathbb{R}^{T\times T}$ is a mask matrix whose element $I_{t_it_j}$ is either $-\infty$ or $0$: $I_{t_it_j}=0$ if $a\leq t_i\leq t_j\leq b$; otherwise, $I_{t_it_j}=-\infty$. Such a mask matrix is used to capture the sequential order of items within a session. The sequential self-attention mechanism produces a zero attention weight, i.e., $\beta^{t_it_j}_{\mathcal{S}^{u}_{a:b}}=0$, if a future $t_i$\textsuperscript{th} item attempts to attend to its past $t_j$\textsuperscript{th} item, i.e., $t_i > t_j$. That said, the matrix of sequential self-attention $\mathbf{\beta}_{\mathcal{S}^{u}_{a:b}}$ is a triangle matrix, whose entry represents the contribution of the $t_i$\textsuperscript{th} item on its following $t_j$\textsuperscript{th} item ($t_i\leq t_j$). Higher attention weights $\beta^{t_it_j}_{\mathcal{S}^{u}_{a:b}}$ indicate that the $t_i$\textsuperscript{th} item has stronger impact on the $t_j$\textsuperscript{th} one.

\subsection{Final Embedding Generation}
We aim to generate the final user and item embeddings for user $u$ and every item $v\in\mathcal{S}^u_{a:b}$, denoted by $\tilde{\mathbf{u}}$ and $\tilde{\mathbf{v}}$, then accordingly perform the sequential recommendation. Although the proposed relational attentive GNN and sequential self-attention are enough to produce final embeddings, we want the embeddings to encode sequential user preference and item correlation in a more fine-grained manner. Hence, we divide every session into $\pi$ subsessions in order, $\mathcal{S}^{u}_{a:b}=\{ \mathcal{S}^{u}_{a:a+\tau-1}, \mathcal{S}^{u}_{a+\tau:a+2\tau-1},..., \mathcal{S}^{u}_{a+(\pi-1)\tau:b}\}$, where $\tau=(a-b+1)/\pi$. We term the complete session $\mathcal{S}^{u}_{a:b}$ as \textit{long-term} item sequence, and each subsession $\mathcal{S}^{u}_{a+(j-1)\tau:a+j\tau-1}$ ($j\in \{ 1,...,\pi\}$) as \textit{short-term} item sequence, denoted by $s^{u}_{j}$ for simplicity. A smaller $\tau$ value leads to more fine-grained short-term sequences. We will discuss how $\tau$ affects the performance in the experiments.

We consider both long-term and short-term sequential information, $\mathcal{S}^u_{a:b}$ and $s^u_{j}$, into the generation of final user and item embeddings. Specifically, for each user $u$ and item $v$, we can generate their corresponding long-term and short-term user and item embeddings by: 
\begin{equation}
\begin{split}
\mathbf{u}[L] &= RAG(\mathcal{H}^h_{a:b}[u,v])[u],\\
\mathbf{v}[L] &= SSA(RAG(\mathcal{H}^h_{a:b}[u,v])[v]),\\
\mathbf{u}_j[S] &= RAG(\mathcal{H}^h_{s_j}[u,v])[u], \forall j\in \{ 1,2,...,\pi\},\\
\mathbf{v}_j[S] &= SSA(RAG(\mathcal{H}^h_{s_j}[u,v])[v]), \forall j\in \{ 1,2,...,\pi\},
\end{split}
\end{equation}
where $[L]$ and $[S]$ denote long-term and short-term representations, respectively, $[u]$ and $[v]$ are used to retrieve the user and item parts, respectively, and functions $RAG$ and $SSA$ represent relational attentive GNN and sequential self-attention, respectively. Here we allow an option to set different numbers of layers, denoted as $L_{lo}$ and $L_{sh}$, for long-term and short-term RA-GNNs, respectively. We will discuss $L_{lo}$ and $L_{sh}$ in the experiment. Last, we utilize the concatenation operator $\oplus$ to combine all of the derived embeddings, and perform an one-hidden-layer feed forward network ($FFN$) to produce the final user and item embeddings $\tilde{\mathbf{u}}$ and $\tilde{\mathbf{v}}$, given by:
\begin{equation}
\begin{split}
\tilde{\mathbf{u}} &= FFN\left(\left[ \mathbf{u}[L] \oplus \mathbf{u}_1[S] \oplus \mathbf{u}_2[S] \oplus ... \oplus \mathbf{u}_\pi[S] \right]\right),\\
\tilde{\mathbf{v}} &= FFN\left(\left[ \mathbf{v}[L] \oplus \mathbf{v}_1[S] \oplus \mathbf{v}_2[S] \oplus ... \oplus \mathbf{v}_\pi[S] \right]\right),
\end{split}
\end{equation}
in which the dimensions of both embeddings $\tilde{\mathbf{u}}$ and $\tilde{\mathbf{v}}$ are $d$.

\subsection{Prediction \& Model Training}
The prediction of the next items consists of two parts. First, we adopt the conventional matrix factorization~\cite{mfhe16} to capture global user interests on items. We perform dot product between user embedding obtained from RA-GNN and primitive item embedding, i.e., $\tilde{\mathbf{u}}$ and $\mathbf{x}_v$, for matrix factorization. 
Second, we incorporate the correlation between existing sequential items and the target item in the prediction. The joint long- and short-term sequential item embedding $\tilde{\mathbf{v}}$ is used here. Given session $\mathcal{S}^u_{a:b}$ created by user $u$, the corresponding predicted score $\hat{y}_{uv}$ on target item $v$ can be generated by:
\begin{equation}
\hat{y}_{uv}=\tilde{\mathbf{u}}\cdot\mathbf{x}_v + \sum_{v_i\in \mathcal{S}^u_{a:b}}\tilde{\mathbf{v}}_i\cdot\mathbf{x}_v,
\end{equation}
where $\mathbf{x}_v\in\mathbf{X}$ is the primitive embedding of item $v$. We expect that the true item $v$ adopted by user $u$ can lead to higher score $\hat{y}_{uv}$.

The overall loss function consists of two main parts. One is the loss for user-item prediction in sequential recommendation (SR), and the other is the \textit{relation-aware regularization} (RAR). We optimize the SR part by Bayesian Personalized Ranking (BPR) objective~\cite{bpr09}, i.e., the pairwise ranking between positive and non-interacted items. We optimize the RAR part by encouraging those embeddings of users, items, and attributes connected by the same relations, along the layers of RA-GNN (from the $l$\textsuperscript{th} to $(l+1)$\textsuperscript{th} layer, to be similar with each other. The loss function is as follows:
\begin{equation}
\label{eq-loss}
\begin{split}
\mathcal{J}=&\mathcal{J}_{SR}+\lambda\mathcal{J}_{RAR}+\eta\|\Theta\|^2_F \\
=&\sum_{(u,v,\mathcal{S}^{u}_{a:b},v')\in\mathcal{D}}-\log\sigma(\hat{y}_{uv}-\hat{y}_{uv'}) \\
&+\lambda\left(\sum_{r\in\mathcal{R}}\sum_{l=1}^{\mathcal{L}-1} \|\mathbf{W}^{l+1}_r-\mathbf{W}^{l}_r\|^{2}_F \right) +\eta\|\Theta\|^2_F
\end{split}
\end{equation}
where $\mathcal{S}^{u}_{a:b}$ denotes a session of user $u$, $v'$ is a non-interacted item (i.e., negative sample), $\mathcal{D}$ is the entire training set, $\mathcal{L}$ is the number of RA-GNN layers, is  $\Theta$ contains all learnable parameters in the neural network, $\lambda$ is the weighting hyperparameter on the RAR term, $\eta$ is the weight for general parameter regularization, and $\|\cdot\|^2_F$ denotes the Frobenius norm of a matrix. The RAR term will restrain the parameter matrices of adjacent RA-GNN layers with the same relation from having too much difference. That said, such a kind of regularization can not only preserve the relation knowledge during RA-GNN training, but also help the optimization distinguish various relations from one another. We will discuss the effect of $\lambda$ on the RAR term in the experiment.
We utilize Adam~\cite{adam15} to adaptively adjust the learning rate during learning. 


\section{experiments} 
\label{sec:experiments}
\begin{table}[!t]
\centering
\caption{Statistics of three datasets. ``n/a'' means that we do not have the information.}
\vspace{-5pt}
\label{tab:data-stat}
\resizebox{\linewidth}{!}{%
\begin{tabular}{c|c|c|c|c}
\hline
 & \# Users & \# Items & \# Interactions & \# Attributes \\ \hline
Instagram & 7,943 & 4,687 & 215,927 & 32\\ 
MovieLens & 1,204 & 3,952 & 125,112 & 20 \\ 
Book-Crossing & 52,406 & 41,264 & 1,856,747 & n/a \\ \hline
\end{tabular}%
}
\end{table}

We conduct experiments to answer three major evaluation questions. (a) How does the proposed RetaGNN perform in conventional, inductive, and transferable sequential recommendation, comparing to the state-of-the-arts and baselines? (b) Does each component in RetaGNN truly take effect? And which component in RetaGNN contributes most to the performance? (c) How do various hyperparameters affect the performance of RetaGNN? 

\subsection{Evaluation Setup}
\textbf{Datasets.} 
Three datasets are employed in the evaluation. (a) \textbf{Instagram} (IG): a user check-in records on locations~\cite{tagv18} at three major urban areas, New York, Los Angeles, and London crawled via Instagram API in 2015. Check-in locations are treated as items. (b) \textbf{MovieLens-1M}\footnote{\url{https://grouplens.org/datasets/movielens/1m/}} (ML): a widely-used benchmark dataset for movie recommendation. (c) \textbf{Book-Crossing}\footnote{\url{http://www2.informatik.uni-freiburg.de/~cziegler/BX/}} (BC): it contains explicit ratings (from $0$ to $10$) of books in the Book-Crossing community. Since MovieLens-1M and Book-Crossing are explicit feedback data, we follow MKR~\cite{mkr19} to convert them into implicit feedback (i.e., $1$ indicates that the user has rated the item and otherwise $0$). The threshold of positive ratings is $4$ for MovieLens-1M $9$ for Book-Crossing. We preprocess the datasets by removing users without any attributes and users containing fewer than $4$ interactions with items. The data statistics is summarized in Table~\ref{tab:data-stat}. 

\begin{table*}[!t]
\centering
\caption{Main results in Precision, Recall, and NDCG for Conventional SR (CSR) over three datasets. The best and second-best performed methods in each metric are highlighted in ``\textbf{bold}'' and \underline{underline}, respectively. The performance improvement is derived by $\frac{RetaGNN-HGAN}{HGAN}\times 100\%$.}
\vspace{-6pt}
\label{tab:exp-main}
\begin{tabular}{c|c|c|c|c|c|c|c|c|c}
\hline
 & \multicolumn{3}{c|}{MoveLens-1M} & \multicolumn{3}{c|}{Book-Crossing} & \multicolumn{3}{c}{Instagram} \\ \hline
 & P@10 & N@10 & R@10  & P@10 & N@10 & R@10  & P@10 & N@10 & R@10 \\ \hline
HGN & 0.1273 & 	0.1368 & 	0.1194 & 	0.0367 & 	0.0392 & 	0.0268 & 	0.0492 & 	0.0516 & 	0.0426  \\ 
SASRec & 0.1173 & 	0.1272 & 	0.1367 & 	0.0272 & 	0.0363 & 	0.0216 & 	0.0378 & 	0.0347 & 	0.0414 \\ 
HAM & 0.1432 & 0.1354 &	0.1447 & 0.0326 & 0.0270 & 0.0628 & 0.0427 & 0.0425 & 0.0425  \\ 
MA-GNN & 0.1536 & 0.1497 & 0.1485 & 0.0512 & 0.0281 & 0.0454 & 0.0514 & 0.0534 & 0.0518  \\ 
HGAN & \underline{0.1614} & \underline{0.1594} & \underline{0.1703} & \underline{0.0535} & \underline{0.0502} & \underline{0.1493} & \underline{0.0594} & \underline{0.0619} & \underline{0.0579}  \\ 
GCMC & 0.1497 & 0.1432 & 0.1573 & 0.0453 & 0.0387 & 0.1432 & 0.0527 & 0.0554 & 0.0418 \\ 
IGMC & 0.1523 & 0.1488 & 0.1531 & 0.0482 & 0.0464 & 0.1356 & 0.0532 & 0.0543 & 0.0441 \\ \hline
RetaGNN & \textbf{0.1704} & \textbf{0.1691} & \textbf{0.1825} & \textbf{0.0693} & \textbf{0.0642} & \textbf{0.1555} & \textbf{0.0692} & \textbf{0.0703} & \textbf{0.0673} \\ \hline
Improvement & 5.58\% & 6.08\% & 7.16\% & 29.53\% & 27.89\% & 4.15\% & 16.50\% & 13.57\% & 16.23\% \\ \hline
\end{tabular}%
\end{table*}

\begin{table*}[!t]
\centering
\caption{Results on Inductive SR (ISR). Percentage values in the first column mean the percentages of users utilized for training. 
}
\vspace{-6pt}
\label{tab:exp-inductive}
\begin{tabular}{c|c|c|c|c|c|c|c|c|c|c}
\hline
 &  & \multicolumn{3}{c|}{MoiveLens} & \multicolumn{3}{c|}{Instagram} & \multicolumn{3}{c}{Book-Crossing} \\ \hline
 &  & P@10 & R@10 & N@10 & P@10 & R@10 & N@10 & P@10 & R@10 & N@10 \\ \hline
\multirow{3}{*}{70\%} & GCMC & 0.1173 & 0.1482 & 0.1326 & 0.0293 & 0.0383 & 0.0425 & 0.0419 & 0.1245 & 0.0225 \\ 
 & IGMC & 0.1238 & 0.1284 & 0.1137 & 0.0356 & 0.0396 & 0.0432 & 0.0425 & 0.1295 & 0.0316 \\ 
 & RetaGNN & \textbf{0.1509} & \textbf{0.1535} & \textbf{0.1515} & \textbf{0.0524} & \textbf{0.0529} & \textbf{0.0573} & \textbf{0.0589} & \textbf{0.1356} & \textbf{0.0545} \\ \hline
\multirow{3}{*}{50\%} & GCMC & 0.1047 & 0.1321 & 0.1032 & 0.0357 & 0.0374 & 0.0407 & 0.0254 & 0.1219 & 0.0236 \\ 
 & IGMC & 0.1145 & 0.1339 & 0.1074 & 0.0362 & 0.0415 & 0.0426 & 0.0395 & \textbf{0.1367} & 0.0378 \\ 
 & RetaGNN & \textbf{0.1356} & \textbf{0.1516} & \textbf{0.1356} & \textbf{0.0568} & \textbf{0.0513} & \textbf{0.0562} & \textbf{0.0585} & 0.1288 & \textbf{0.0528} \\ \hline
\multirow{3}{*}{30\%} & GCMC & 0.0821 & 0.1021 & 0.0832 & 0.0305 & 0.0306 & 0.0386 & 0.0215 & 0.0795 & 0.0228 \\ 
 & IGMC & 0.1028 & 0.1242 & 0.0942 & 0.0318 & 0.0421 & 0.0391 & 0.0389 & 0.0895 & 0.0342 \\ 
 & RetaGNN & \textbf{0.1324} & \textbf{0.1494} & \textbf{0.1245} & \textbf{0.0525} & \textbf{0.0494} & \textbf{0.0473} & \textbf{0.0563} & \textbf{0.1053} & \textbf{0.0494} \\ \hline

\end{tabular}%
\end{table*}

\textbf{Competing Methods.}
We compare the proposed RetaGNN with several SOTA methods and baselines. Their settings of hyperparameters are tuned by grid search on the validation set. 

\begin{itemize}[leftmargin=*]
\item \textbf{SASRec}~\footnote{\url{https://github.com/kang205/SASRec}}~\cite{sasrec18}: a self-attention based sequential model that utilizes the attention mechanism to identify relevant items and their correlation in entire item sequences.
\item \textbf{HGN}~\footnote{\url{https://github.com/allenjack/HGN}}~\cite{hgn19}: a hierarchical gating network that learns the item subsequence embeddings through feature gating in long and short aspects, and models the item-item relations.
\item \textbf{HAM}~\cite{peng2020ham}~\footnote{\url{https://github.com/BoPeng112/HAM}} (SOTA): a hybrid association model with simplistic pooling to generate sequential recommendations based on users' long-term preferences, high- and low-order sequential association patterns between users and items.
\item \textbf{MA-GNN}~\cite{ma2020memory} (SOTA): a memory-augmented graph neural network to model the short-term contextual information of items, together with a shared memory network to encode the long-range dependencies between items for sequential recommendation.
\item \textbf{HGAN}~\cite{wang2019heterogeneous}~\footnote{\url{https://github.com/Jhy1993/HAN}} (SOTA): a heterogeneous graph attention network model that generate node embeddings by aggregating features from meta-path based neighbors in a hierarchical manner. We utilize HGAN to produce the embeddings of users and items in the constructed tripartite graph.
\item \textbf{GCMC}~\cite{berg2017graph}~\footnote{\url{https://github.com/riannevdberg/gc-mc}}: a graph autoencoder framework that produces user and item embeddings through differentiable message passing on the bipartite interaction graph, along with a bilinear decoder to predict user-item interactions. GCMC can be used for both inductive and transferable RS.
\item \textbf{IGMC}~\cite{igmc}~\footnote{\url{https://github.com/muhanzhang/IGMC}} (SOTA): an inductive graph neural network model that maps the subgraph extracted from 1-hop neighborhood of a user-item pair to its corresponding rating. IGMC can be used for both inductive and transferable RS.
\end{itemize}

Note that the first four competitors, including SASRec, HGN, HAM, and MA-GNN, are recent advanced models for sequential recommendation. The last three, i.e., HGAN, GCMC, and IGMC, are recent powerful graph neural network-based models for general recommendation systems, in which IGMC is the first GNN-based model on being capable of inductive and transfer learning.

\textbf{Evaluation Metrics.}
Three evaluation metrics are utilized for performance comparison: Precision@k (P@$k$), Recall@k (R@$k$), and NDCG@k (N@$k$). P@$k$ is used to estimate whether a method can find the ground-truth items for a user when the first $k$ recommended items are reported.
R@$k$ indicates what percentage of a user's interacted items can emerge in the top $k$ recommended items. N@$k$ is the normalized discounted cumulative gain at top $k$ positions, which takes the positions of correctly recommended items into account. We fix $k=10$ throughout the experiments since other $k$ values exhibit similar results.

\textbf{Evaluation Settings.}
The ratio of training, validation, and test sets is 6:2:2. We repeat every experiment $10$ times, and report the average results. We fix the session length $t=11$ and the future length (i.e., number of future items to be predicted) $g=3$, by default, and will report the results of varying $t$ and $g$. In RetaGNN, we set the weighting factor of relation-aware regularization term as $\lambda=0.6$.  
The hop number to extract $h$-hop enclosing subgraphs is set as $h=2$ by default. We set the dimensionality of every embedding $d=32$.
The default numbers of long-term and short-term RA-GNN layers are $L_{lo}=2$ and $L_{sh}=3$.
The length of each short-term subsession is set as $\tau=4$ by default.
We examine how different hyperparameters affect the performance in the following. All experiments are conducted with PyTorch running on GPU machines (Nvidia GeForce GTX 1080 Ti). The Adam optimizer is set with an initial learning rate $0.001$ and the batch size $32$.

\subsection{Main Experimental Results}
\label{sec-expres}

\textbf{Results on CSR.} The main results on conventional SR are shown in Table~\ref{tab:exp-main}. We can find that the proposed RetaGNN significantly outperforms all of the competing methods on all metrics in the task of CSR. We can draw four findings based on the results. First, RetaGNN outperforms the four state-of-the-art SR methods. It implies that modeling the sequential high-order interactions between users and items, which is neglected in these competing methods, does take effect. Second, the recent advance of heterogeneous graph neural network HGAN is the second-best model. Since HGAN does not capture the sequential patterns of user preferences in the given session, it performs worse than our RetaGNN. Third, although IGMC and GCMC can learn the high-order interactions between users and items, their embedding generation methods cannot model the dynamics of user-item interactions from the given sequence. Hence, the performance of IGMC and GCMC is still worse than RetaGNN. 
Fourth, RetaGNN outperforms baselines better on Book-Crossing and Instagram than on MovieLens. We think the reason is about data sparsity, i.e., the density of user-item interactions. We can define ``interaction density'' as: \#interactions/(\#users $times$ \#items). Then the density values of MovieLens, Instagram, and Book-Crossing are $0.0263$, $0.0058$, and $0.0009$, respectively. It is apparent that higher density leads to less performance improvement of RetaGNN. Dense user-item interactions make baselines easier to learn user preferences. RetaGNN is verified to better generate user representations based on sparse interactions.
In a nutshell, RetaGNN brings a key insight: it is crucial for a CSR model to incorporate the high-order user-item interactions and sequential patterns of user preferences into the representation learning of users and items.

\begin{table}[!t]
\centering
\caption{Results on Transferable SR (TSR). In ``A $\rightarrow$ B'', A and B are the source and target domain, indicating using data in A domain to train the model, and predicing on B domain.}
\vspace{-5pt}
\label{tab:exp-transfer}
\resizebox{\linewidth}{!}{%
\begin{tabular}{c|c|c|c|c|c|c}
\hline
 & \multicolumn{3}{c|}{ML $\rightarrow$ BC} & \multicolumn{3}{c}{ML $\rightarrow$ IG} \\ \hline
 & P@10 & N@10 & R@10  & P@10 & N@10 & R@10 \\ \hline
GCMC & 0.0343 & 	0.0326 & 	0.1178 & 	0.0439 & 	0.0450 & 	0.0334\\ 
IGMC & 0.0397 & 	0.0364 & 	0.1297 & 	0.0449 & 	0.0438 & 	0.0372 \\ 
RetaGNN & \textbf{0.0502} & 	\textbf{0.0595} & 	\textbf{0.1374} & 	\textbf{0.0599} & 	\textbf{0.0608} & 	\textbf{0.0596} \\ \hline \hline
 & \multicolumn{3}{c|}{BC $\rightarrow$ ML} & \multicolumn{3}{c}{BC $\rightarrow$ IG} \\ \hline
 & P@10 & N@10 & R@10  & P@10 & N@10 & R@10 \\ \hline
GCMC & 0.1193 & 	0.1056 & 	0.1245 & 	0.0394 & 	0.0381 & 	0.0407 \\ 
IGMC & 0.1294 & 	0.1146 & 	0.1321 & 	0.0381 & 	0.0402 & 	0.0413 \\ 
RetaGNN & \textbf{0.1456} & 	\textbf{0.1398} & 	\textbf{0.1468} & 	\textbf{0.0643} & 	\textbf{0.0605} & 	\textbf{0.0585} \\ \hline \hline
 & \multicolumn{3}{c|}{IG $\rightarrow$ ML} & \multicolumn{3}{c}{IG $\rightarrow$ BC} \\ \hline
 & P@10 & N@10 & R@10  & P@10 & N@10 & R@10 \\ \hline
GCMC & 0.1377 & 	0.1312 & 	0.1453 & 	0.0385 & 	0.0301 & 	0.1256 \\ 
IGMC & 0.1473 & 	0.1373 & 	0.1583 & 	0.0413 & 	0.0327 & 	0.1274 \\ 
RetaGNN & \textbf{0.1684} & 	\textbf{0.1606} & 	\textbf{0.1742} & 	\textbf{0.0604} & 	\textbf{0.0599} & 	\textbf{0.1452} \\ \hline
\end{tabular}%
}
\end{table}

\begin{table*}[!t]
\centering
\caption{Results on ablation studies in CSR by removing one of RetaGNN components. 
}
\vspace{-6pt}
\label{tab:exp-ablation}
\begin{tabular}{c|c|c|c|c|c|c|c|c|c}
\hline
 & \multicolumn{3}{c|}{MoveLens-1M} & \multicolumn{3}{c|}{Book-Crossing} & \multicolumn{3}{c}{Instagram} \\ \hline
 & P@10 & N@10 & R@10  & P@10 & N@10 & R@10  & P@10 & N@10 & R@10 \\ \hline
- RA-GNN & 0.1368 &	0.1383 & 	0.1425 & 	0.0384 & 	0.0427 & 	0.1267 & 	0.0507 & 	0.0519 & 	0.0437 \\
- Attribute nodes & 0.1593 & 0.1572 & 0.1704 & 0.0492 & 0.0485 & 0.1399 & 0.0604 & 0.0596 & 0.0554 \\ 
- RAR term in loss & 0.1583 & 0.1589 & 0.1715 & 0.0468 & 0.0548 & 0.1397 & 0.0571 & 0.0572 & 0.0488  \\ 
- Relational Attention & 0.1576 & 0.1528 & 0.1636 & 0.0405 & 0.0495 & 0.1405 & 0.0521  & 0.0542 & 0.0448  \\ 
- Sequential Self-Attention & 0.1613 & 0.1625 & 0.1734 & 0.0536 & 0.0593 & 0.1495 & 0.0610 & 0.0628 & 0.0575 \\ 
- Short-term part & 0.1627 & 0.1642 & 0.1782 & 0.0538 & 0.0604 & 0.1487 & 0.0620 & 0.0630 & 0.0587 \\ 
- Long-term part & 0.1635 & 0.1592 & 0.1723 & 0.0524 & 0.0626 & 0.1413 & 0.0567 & 0.0642 & 0.0525 \\ \hline
Full RetaGNN & 0.1704 & 0.1691 & 0.1825 & 0.0693 & 0.0642 & 0.1555 & 0.0692 & 0.0703 & 0.0673 \\ \hline
\end{tabular}%
\end{table*}

\textbf{Results on ISR.} To conduct the ISR experiments, we randomly divide the entire user set $\mathcal{U}$ in each dataset into two subsets $\mathcal{U}^{train}$ and $\mathcal{U}^{test}$, and ensure $\mathcal{U}^{train} \cap \mathcal{U}^{test} = \emptyset$. Users in $\mathcal{U}^{test}$ are regarded as the new-coming ones, whose embeddings cannot be obtained at the training stage. By varying the percentage of users in $\mathcal{U}^{train}$ as 70\%, 50\% to 30\% (users in $\mathcal{U}^{test}$ are 30\%, 50\%, and 70\% correspondingly), we present the results on inductive SR in Table~\ref{tab:exp-inductive}. Note that we compare RetaGNN with only the state-of-the-art inductive recommendation models, GCMC and IGMC, because other methods are not applicable in the inductive setting. The results exhibit that RetaGNN can significantly outperform IGMC and GCMC in all metrics over three datasets. As the training user percentage decreases (e.g., to 30\%), the superiority of RetaGNN keeps significant. Although IGMC is also a graph neural network-based recommendation model, which similarly learns the embeddings from the extracted subgraph surrounded by the target user-item pair, it models neither the fine-grained (long-term and short-term) dynamics of sequential items nor the relation-aware regularization, which bring the outstanding inductive capability of RetaGNN.

\textbf{Results on TSR.} Since we have three datasets that come from diverse domains, to conduct the experiments of transferable SR, we consider each dataset as the source-domain data for training, and treat the remaining two datasets as the target-domain data for testing. There are six ``source $\rightarrow$ target'' data combinations: ML $\rightarrow$ BC, ML $\rightarrow$ IG, BC $\rightarrow$ ML, BC $\rightarrow$ IG, IG $\rightarrow$ ML, and IG $\rightarrow$ BC. It is clear that both sets of users and items in source and target domains are disjoint. Since the Book-Crossing data does not contain item attributes, the TSR experiments consider only user-item bipartite graphs for the learning of graph neural networks. The results on TSR are displayed in Table~\ref{tab:exp-transfer}. We compare RetaGNN with IGMC and GCMC because they are state-of-the-art transfer learning-based recommenders without relying on content and auxiliary information. It can be obviously observed that RetaGNN again leads to the best performance, which is significantly better than IGMC and GCMC, in all of the six source-target data combinations. Such results prove the usefulness of transferring relational attentive weight matrices across domains under the graph neural network framework, and the relation-aware regularization in RetaGNN.

\begin{table*}[!t]
\centering
\caption{The effect of the training session length $t$ and the number of next predicted items $g$ on CSR with RetaGNN.}
\vspace{-6pt}
\label{tab:exp-length}
\begin{tabular}{c|c|c|c|c|c|c|c|c|c}
\hline
 & \multicolumn{3}{c|}{Instagram} & \multicolumn{3}{c|}{MovieLens} & \multicolumn{3}{c}{Book-Crossing} \\ \hline
 & P@10 & N@10 & R@10 & P@10 & N@10 & R@10 & P@10 & N@10 & R@10 \\ \hline
$t$=9, $g$=1 & 0.1592 & 	0.1599 & 	0.1702 & 	0.0512 & 	0.0569 & 	0.1414 & 	0.0557 & 	0.0573 & 	0.0547 \\
$t$=9, $g$=2 & 0.1604 & 	0.1616 & 	0.1757 & 	0.0535 & 	0.0586 & 	0.1432 & 	0.0583 & 	0.0596 & 	0.0561 \\
$t$=9, $g$=3 & 0.1623 & 	0.1629 & 	0.1765 & 	0.0598 & 	0.0599 & 	0.1474 & 	0.0607 & 	0.0604 & 	0.0584 \\
$t$=11, $g$=1 & 0.1656 & 	0.1627 & 	0.1714 & 	0.0627 & 	0.0595 & 	0.1442 & 	0.0604 & 	0.0636 & 	0.0603 \\
$t$=11, $g$=2 & 0.1668 & 	0.1686 & 	0.1752 & 	0.0652 & 	0.0613 & 	0.1435 & 	0.0635  &	0.0674 & 	0.0631 \\
$t$=11, $g$=3 & \textbf{0.1704} & 	\textbf{0.1691} & 	\textbf{0.1825} & 	\textbf{0.0693} & 	\textbf{0.0642} & 	\textbf{0.1555} &  	\textbf{0.0692}  &	\textbf{0.0703} & 	\textbf{0.0673} \\
$t$=13, $g$=1 & 0.1654 & 	0.1636 & 	0.1763 & 	0.0536  &	0.0598 & 	0.1483  &	0.0615 & 	0.0652 & 	0.0604 \\
$t$=13, $g$=2 & 0.1669 & 	0.1658 & 	0.1774 & 	0.0593 & 	0.0614 & 	0.1475 & 	0.0639 & 	0.0616 & 	0.0649 \\
$t$=13, $g$=3 & 0.1683 & 	0.1667 & 	0.1785 & 	0.0634 & 	0.0625  &	0.1508 & 	0.0656 &  	0.0594 & 	0.0625\\ \hline
\end{tabular}%
\end{table*}

\begin{figure}[!t]
  \centering
  \includegraphics[width=1.0\linewidth]{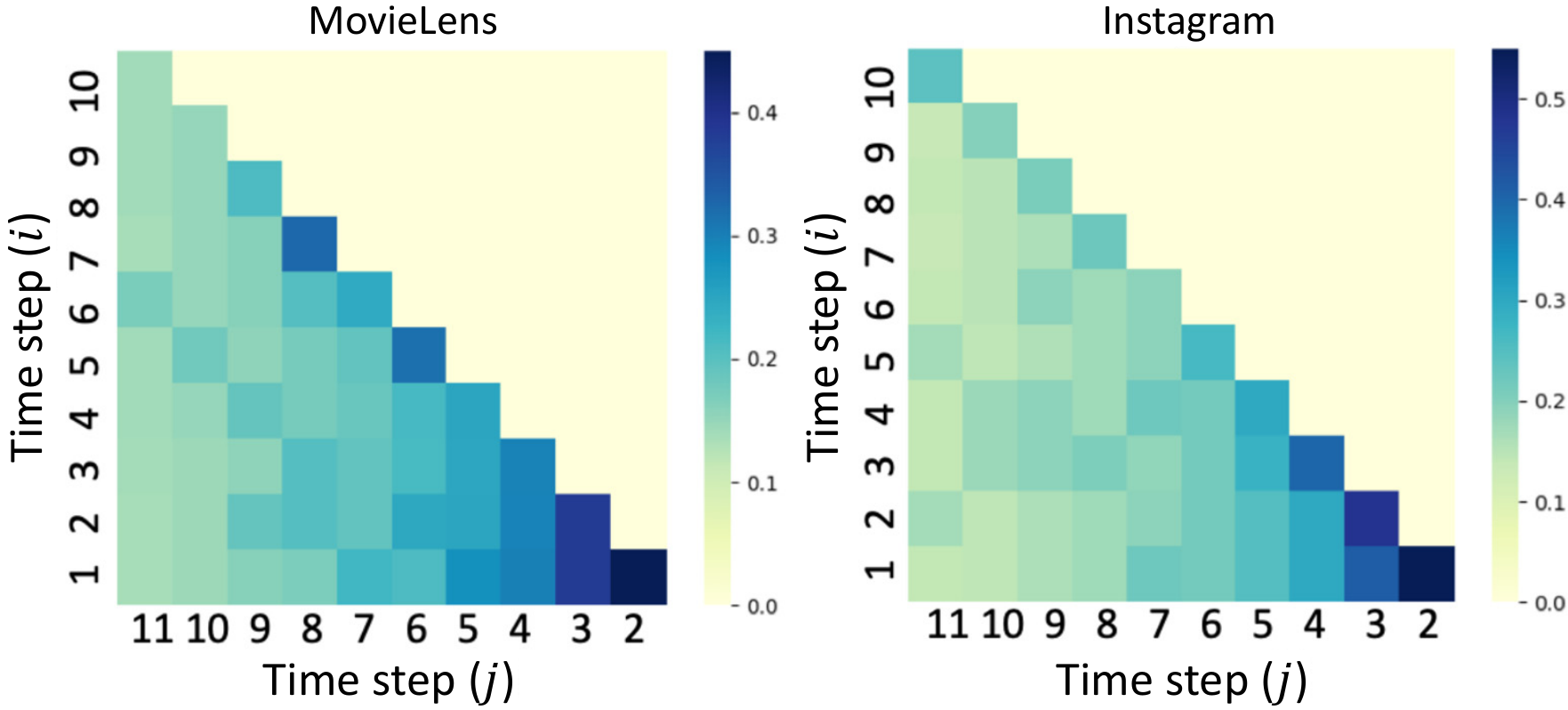}
  \caption{Heatmap of mean sequential self-attention weights trained for CSR on MovieLens and Instagram datasets. Attention weights tend to be uniformly distributed in MovieLens, while Instagram exhibits a bias towards recent time steps.}
  \label{fig:exp-attvis}
\end{figure}

\begin{table*}[!t]
\centering
\caption{The CSR performance of hop number $h$ for $h$-hop enclosing subgraphs, and window size $\tau$ of short-term subsessions.}
\vspace{-6pt}
\label{tab:exp-hop-winow}
\begin{tabular}{c|c|c|c|c|c|c|c|c|c}
\hline
 & \multicolumn{3}{c|}{Instagram} & \multicolumn{3}{c|}{MovieLens} & \multicolumn{3}{c}{Book-Crossing} \\ \hline
 & P@10 & N@10 & R@10 & P@10 & N@10 & R@10 & P@10 & N@10 & R@10 \\ \hline
$h$=2, $\tau$=3 & 0.1684 & 	0.1643 & 	0.1792 & 	0.0676 & 	\textbf{0.0663} & 	0.1496 & 	0.0673 & 	0.0691 & 	0.0669 \\ 
$h$=2, $\tau$=4 & \textbf{0.1704} & 	0.1691 & 	\textbf{0.1825} & 	\textbf{0.0693} & 	0.0642 & 	\textbf{0.1555} & 	\textbf{0.0692} & 	\textbf{0.0703} & 	\textbf{0.0673} \\ 
$h$=2, $\tau$=5 & 0.1696 & 	\textbf{0.1718} & 	0.1819 & 	0.0684 & 	0.0657 & 	0.1484 & 	0.0684 & 	0.0685 & 	0.0684 \\ 
$h$=3, $\tau$=3 & 0.1668 & 	0.1627 & 	0.1768 & 	0.0662 & 	0.0625 & 	0.1478 & 	0.0652 & 	0.0667 & 	0.0653 \\ 
$h$=3, $\tau$=4 & 0.1674 & 	0.1653 & 	0.1772 & 	0.0672 & 	0.0636 & 	0.1472 & 	0.0665 & 	0.0673 & 	0.0648 \\ 
$h$=3, $\tau$=5 & 0.1671 & 	0.1669 & 	0.1793 & 	0.0679 & 	0.0648 & 	0.1467 & 	0.0667  &	0.0652 & 	0.0659 \\ 
$h$=4, $\tau$=3 & 0.1646 & 	0.1618 & 	0.1736 & 	0.0647 & 	0.0608 & 	0.1453 & 	0.0645 & 	0.0635 & 	0.0645 \\ 
$h$=4, $\tau$=4 & 0.1653 & 	0.1627 & 	0.1757 & 	0.0658 & 	0.0614 & 	0.1462 & 	0.0651 & 	0.0631 & 	0.0636 \\ 
$h$=4, $\tau$=5 & 0.1667 & 	0.1639 & 	0.1773 & 	0.0673 & 	0.0629 & 	0.1469 & 	0.0642 & 	0.0638 & 	0.0647 \\ \hline
\end{tabular}%
\end{table*}

\subsection{Model Analysis}
\label{sec-modelanalysis}

\textbf{Ablation Study.}
To understand how each component in RetaGNN contributes to the overall performance, we conduct the ablation study. We compare the full RetaGNN model with seven of its variants that remove one component. The seven variants include: (a) removing the RA-GNN layer, (b) model without attributes (i.e., using only user-item bipartite graph), (c) removing the RAR term in the loss function, (d) removing the relational attention $\alpha^l_{ij}$ in RA-GNN (i.e., message passing by only learnable $\mathbf{W}^l_o$ and $\mathbf{W}^l_r$), (e) removing the sequential self-attention layer, (f) removing the short-term RA-GNN part in final embedding generation, and (g) removing the long-term RA-GNN part in final embedding generation. 
The ablation analysis is conducted on conventional SR, and the results are exhibited in Table~\ref{tab:exp-ablation}, from which We have obtained four key findings. First, the RA-GNN layer contributes most to the performance of RetaGNN. That is, it is crucial to model the sequential high-order interactions between users and items. Second, RetaGNN without the relational attention $\alpha^l_{ij}$ also leads to a significant performance drop. Such an outcome implies that we do need an attention mechanism to distinguish the importance of various relations in the tripartite graph when learning relational weight matrices. Third, the relation-aware regularization (RAR) has a non-ignorable contribution to RetaGNN. The performance of RetaGNN without RAR can become a bit worse than HGAN presented in Table~\ref{tab:exp-main}. Last, although losing each of sequential self-attention, short-term and long-term parts results in small performance damage, they do bring positive effect for RetaGNN.

\begin{table*}[!t]
\centering
\caption{The CSR performance of layer numbers $L_{lo}$ and $L_{sh}$ for long-term and short-term RA-GNNs, respectively.}
\vspace{-6pt}
 \label{tab:exp-gnnlayer}
\begin{tabular}{c|c|c|c|c|c|c|c|c|c}
\hline
 & \multicolumn{3}{c|}{Instagram} & \multicolumn{3}{c|}{MovieLens-1M} & \multicolumn{3}{c}{Book-Crossing} \\ \hline
 & P@10 & N@10 & R@10 & P@10 & N@10 & R@10 & P@10 & N@10 & R@10 \\ \hline
$L_{lo}$=1, $L_{sh}$=1 & 0.1605 & 	0.1613 & 	0.1774 & 	0.0588 & 	0.0589 & 	0.1463 & 	0.0632 & 	0.0642 & 	0.0636  \\ 
$L_{lo}$=1, $L_{sh}$=2 & 0.1632 & 	0.1639 & 	0.1783 & 	0.0592 & 	0.0584 & 	0.1484 & 	0.0649 & 	0.0659 & 	0.0653  \\ 
$L_{lo}$=1, $L_{sh}$=3 & 0.1659 & 	0.1637 & 	0.1801 & 	0.0624 & 	0.0599 & 	0.1498 & 	0.0635 & 	0.0671 & 	0.0634 \\ 
$L_{lo}$=2, $L_{sh}$=1 & 0.1618 & 	0.1604 & 	0.1795 & 	0.0641 & 	0.0622 & 	0.1493 & 	0.0645 & 	0.0663 & 	0.0646 \\ 
$L_{lo}$=2, $L_{sh}$=2 & 0.1643 & 	0.1649 & 	0.1813 & 	\textbf{0.0701} & 	0.0614 & 	0.1513 & 	0.0653 & 	0.0684 & 	0.0662 \\ 
$L_{lo}$=2, $L_{sh}$=3 & \textbf{0.1704} & 	\textbf{0.1691} & 	\textbf{0.1825} & 	0.0693 & 	\textbf{0.0642} & 	\textbf{0.1555} &	\textbf{0.0692} & 	\textbf{0.0703} & 	0.0673 \\ 
$L_{lo}$=3, $L_{sh}$=1 & 0.1649 & 	0.1642 & 	0.1799 & 	0.0688 & 	0.0629 & 	0.1491 & 	0.0663 & 	0.0653 & 	0.0651  \\ 
$L_{lo}$=3, $L_{sh}$=2 & 0.1638 & 	0.1594 & 	0.1742 & 	0.0653 & 	0.0614 & 	0.1474 & 	0.0651 & 	0.0631 & 	\textbf{0.0689} \\ 
$L_{lo}$=3, $L_{sh}$=3 & 0.1592 & 	0.1583 & 	0.1713 & 	0.0649 & 	0.0593 & 	0.1415 & 	0.0649 & 	0.0625 & 	0.0635 \\ \hline
\end{tabular}%
\end{table*}

\begin{figure*}[!t]
  \centering
  \includegraphics[width=.95\textwidth]{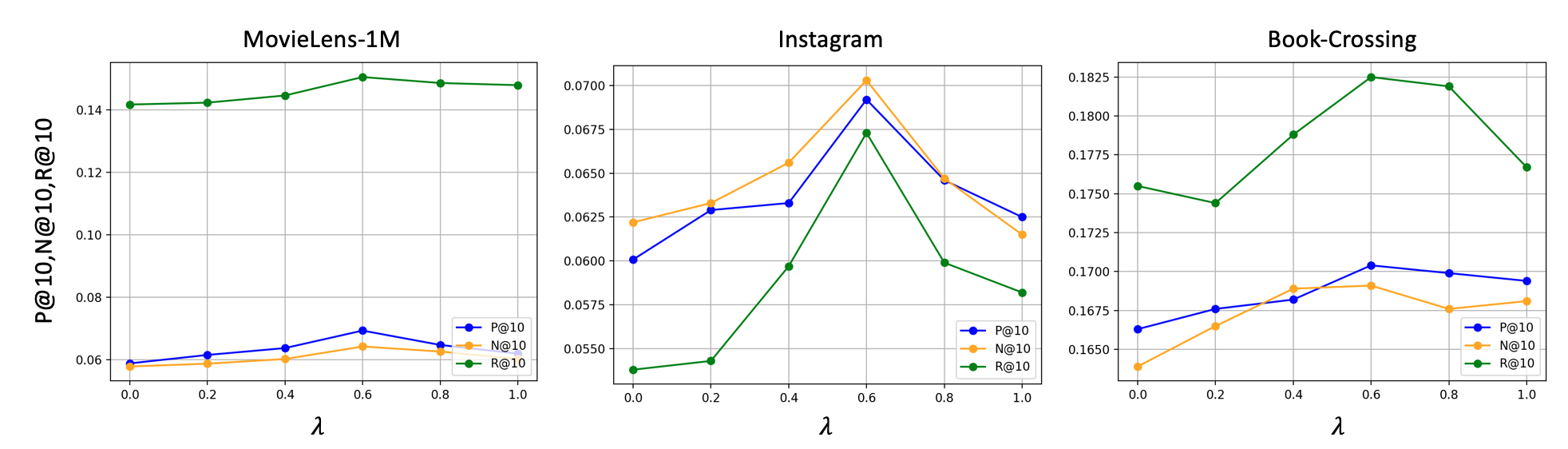}
  \vspace{-5pt}
  \caption{The CSR performance by changing the value of RAR balancing factor $\lambda$.}
  \label{fig:exp-rar}
  \vspace{-5pt}
\end{figure*}

\textbf{Effect of Session Length.}
We examine how the performance of RetaGNN is affected by varying the length of training session $t$ and the number of next items to be predicted $g$. We conduct the analysis on nine combinations of $(t,g)$ pairs with $t\in\{9,11,13\}$ and $g\in\{1,2,3\}$, and show the results in Table~\ref{tab:exp-length}. The length setting with the best performance lies in $(t,g)=(11,3)$, which is the default setting of RetaGNN. 
In general, with fixed $g$, moderately increasing $t$ (e.g., to 11) can improve the performance, but seeing too much historical information (e.g., to $13$) can damage the performance. A potential reason is that longer training sessions (larger $t$) would bring items, which are irrelevant to determine the next items, into model learning. We would suggest using the validation set to determine a proper $t$. On the other hand, it is natural that higher $g$ with fixed $t$ results in better performance because seeing more targeted items provides the model more evidence to learn user preferences.

\textbf{Attention Visualization.}
We visualize the distribution of sequential self-attention weights learned by RetaGNN on MovieLens and Instagram datasets. The visualization plots by heatmap are displayed in Figure~\ref{fig:exp-attvis}, in which attention weights are mean values averaged over all sessions with 11 time steps ($t=11$). Each row represents the sequence of attention weights over the $j$-th historical time steps for the current $i$-th recent time step ($i < j$). Smaller/larger values of $i$ and $j$ refer to newer/older steps. We can find that the attention weights tend to be biased towards recent steps in Instagram. The distribution of attention weights in MovieLens is more uniform. These observations can be connected to real-world intuitions in such two datasets. Since the check-in behaviors on Instagram have a higher potential to be bursty and highly impacted by recent campaigns and promotions. In MovieLens, users tend to follow their preferences to see/rate movies, in which the preferences may keep stable and change slowly over time. These results exhibit that the proposed sequential self-attention can reflect various human behaviors hidden in different domains.

\subsection{Hyperparameter Study}
\label{sec-hyperstudy}

\textbf{Hop Number \& Length of Short-term Subsession.}
By varying the hop number $h\in\{2,3,4\}$, we aim to understand how the extraction of enclosing subgraphs centered at the target user-item pairs affects the performance. We also adjust the length of short-term subsessions $\tau\in\{3,4,5\}$, which determines the granularity of short-term subsessions, to see the performance change. By pairing $(h,\tau)$, we present the results in Table~\ref{tab:exp-hop-winow}, which bring two findings. First, a larger $h$ with fixed $\tau$ leads to worse results. A proper hop number with better performance is $h=2$. The reason should be that larger $h$ brings irrelevant users and items further away from the target pair into the enclosing subgraphs. Noisy enclosing subgraphs can hurt the performance. Second, although $\tau=4$ with fixed $h$ is slightly helpful, the performance improvement of varying $\tau$ is limited. 
Moreover, the average node number of 3-hop neighbors in the tripartite graph is much more than that of 2-hop neighbors. Higher $h$ and $\tau$ could include more noise and result in the over-smoothing problem.
We would suggest to have $(h,\tau)=(2,4)$.

\textbf{Numbers of Long-term and Shor-term RA-GNN Layers.}
We change the numbers of RA-GNN layers in long and short terms, $L_{lo}\in\{1,2,3\}$ and $L_{sh}\in\{1,2,3\}$, to understand how they affect the performance. The results are presented in Table~\ref{tab:exp-gnnlayer}. It can be observed that $(L_{lo},L_{sh})=(2,3)$ leads to better performance. Larger numbers of RA-GNN layers can learn higher-order interactions between users and items, but bring a higher risk of causing over-smoothing. Since the long-term enclosing subgraphs can contain more users, items, and attributes, along with edges connected with them, increasing $L_{lo}$ can be more sensitive to include noise. The short-term enclosing subgraphs represent fine-grained interactions between users and items, and thus are less sensitive to increasing $L_{sh}$. 

\textbf{RAR Weighting Factor $\lambda$.}
The hyperparameter $\lambda$ in the loss function determines the contribution of the relation-aware regularization (RAR) term. We vary $\lambda\in\{0.0,0.2,0.4,0.6,0.8,1.0\}$ to examine whether the performance is sensitive to $\lambda$. The results are displayed in Figure~\ref{fig:exp-rar}, and reveal that $\lambda=0.6$ leads to better performance. We think it is necessary to have a moderate choice of $\lambda$ so that the attentive weight matrices $\mathbf{W}^l_r$ can be mildly retrained across layers. Larger $\lambda$ values could make RA-GNN less flexible to capture high-order user-item interactions.


\section{conclusion and future work} 
\label{sec:conclusion_and_future_work}
In this paper, we propose to solve the holistic sequential recommendation (SR) task, i.e., have a model being capable of conventional, inductive, and transferable SR at the same time. We develop a novel graph neural network (GNN) based model, Relational Temporal Attentive GNN (RetaGNN), to achieve the goal without relying on content and auxiliary information. The inductive and transferable capabilities of RetaGNN come from learning relational attentive weight matrices in the enclosing subgraphs centered at the target user-item pair. The performance improvement of RetaGNN lies in better modeling the sequential high-order interactions between users and items by the RA-GNN layer, and the temporal patterns of user preferences by the sequential self-attention layer. Experiments conducted on three well-known datasets exhibit significantly better performance than state-of-the-arts in holistic SR. A series of evaluation studies robustly verify every design of components and hyperparameters in RetaGNN.

The future extension of RetaGNN is two-fold. First, items can be correlated based on the underlying knowledge graph. We would like to better represent users and items through jointly learning holistic SR and knowledge graph embedding. Second, as RetaGNN is able to perform inductive learning, the SR task can be extended to conversational recommendation. That said, user feedback collected from in chatbot conversation can be instantly thrown into RetaGNN to update user and item embeddings.


\begin{acks}
This work is supported by Ministry of Science and Technology (MOST) of Taiwan under grants 109-2636-E-006-017 (MOST Young Scholar Fellowship) and 109-2221-E-006-173, and also by Academia Sinica under grant AS-TP-107-M05.
\end{acks}

\bibliographystyle{ACM-Reference-Format}
\bibliography{normal_generated}


\end{document}